# Quantitative measurements of the lift force acting on a sphere sliding along a liquid-liquid interface

Hao Zhang[1, 2], Zaicheng Zhang[3], Abdelhamid Maali[2,*]

[1] *State Key Laboratory of Structural Chemistry, Fujian Institute of Research on the Structure of Matter, Chinese Academy of Sciences, Fuzhou 350002, China*
[2] *Université de Bordeaux & CNRS, LOMA, UMR 5798, F-33400 Talence, France*
[3] *School of Physics, Beihang University, 100191 Beijing, China*

**Abstract**

This work explores the lift force experienced by a particle moving in a viscous fluid near a liquid-liquid interface. The lift force is induced by the interaction between the viscous flow generated by the particle's motion and the deformation of the soft interface. The factors influencing the lift force including the velocity, the viscosity, and the sphere radius, and the separation distance were systematically studied. The experiments demonstrate that the lift force intensifies as the particle approaches the interface, and saturates at shorter distances. These findings are consistent with predictions made using soft lubrication theory and numerical calculations, providing strong validation for the theoretical framework.

Keywords: interfacial deformation, elastohydrodynamic (EHD) coupling, lift force, atomic force microscopy (AFM)

* Corresponding author:

Abdelhamid Maali: abdelhamid.maali@u-bordeaux.fr

# 1. Introduction

Lubrication involves introducing a thin fluid layer between two solid surfaces in contact to prevent direct adhesion, minimize wear, and reduce frictional losses. While this fluid reduces direct contact and allows the surfaces to slide more easily against each other, it also creates high pressures in the confined space between them. Indeed, in a classical lubrication scenario with two rigid surfaces, the time-reversibility of the Stokes equations, combined with symmetric boundary conditions, implies that no net lift force can appear. However, in reality, many biologically and technologically relevant systems involve boundaries that are soft or deformable, or objects that are themselves highly compliant. In such cases, the interplay of fluid flow and elastic (or capillary) deformation can break the time-reversal symmetry of the low-Reynolds-number flow. This coupling is often referred to as elastohydrodynamic (EHD) lubrication or soft lubrication, and it generically produces a net normal (lift) force.

The resulting lift force can be understood through a scaling argument that considers the interactions between the fluid dynamics and the elastic deformation of the materials involved. Let us consider the motion of a rigid cylinder with radius $R$, length $L$ completely immersed in a fluid with viscosity $\eta$ and moving along an interface with a velocity $V$, as shown in **Figure 1**. When the substrate is rigid (**Figure 1a**), the viscous pressure $p_0(x)$ between the confining surfaces scales $\eta xV/h^2$, behaving antisymmetric along the $x$ axis, where $h(x)$ is the fluid thickness separating the surfaces [1] (**Figure 1b**). It is positive upstream and negative downstream. Therefore, the integrated normal force is equal to zero: $F_N = L\int_{-\infty}^{+\infty} p(x)dx = 0$. However, when the substrate is a soft solid (**Figure 1c**) with a shear modulus of $G$, the deformation $\delta(x)$ of the substrate causes the pressure profile to lose its antisymmetry (**Figure 1d**) as $p(x) \sim \frac{\eta xV}{(h+\delta)^2} \sim p_0(x) - \frac{2\delta(x)}{h}p_0(x)$ [1]. Given that the deformation of the soft substrate scales as $\delta(x) \sim -\frac{\sqrt{Rh}\,p_0(x)}{G}$ [1], it results in a nonzero normal force at the second leading order, with a scaling law of $F_N \sim \frac{\eta^2 V^2 R^2 L}{Gd^3}$, where $d$ is the gap distance between the apex of cylinder and the undeformed substrate, $G$ is the shear modulus of the material, for $G$ approaching infinity, the normal force $F_N$ vanishes.

When the cylinder is replaced by a spherical particle with a radius of $R$, the scaling law becomes $F_N \sim \frac{\eta^2 V^2}{G}\left(\frac{R}{d}\right)^{\frac{5}{2}}$, where the cylinder length $L$ is replaced by hydrodynamic radius $\sqrt{Rd}$. For the case of a sphere moving along a liquid-liquid interface, where the relation between the liquid-liquid interface deformation and the pressure is $p_0(x) \sim \sigma\frac{\delta(x)}{l^2}$, where $\sigma$ is the surface tension of the liquid-liquid interface and $l$ is the characteristic length of the interaction (i.e. hydrodynamic radius $\sqrt{Rd}$), this results in lift force scaling as: $F_N \sim \frac{\eta^2 V^2 R^3}{\sigma d^2}$.

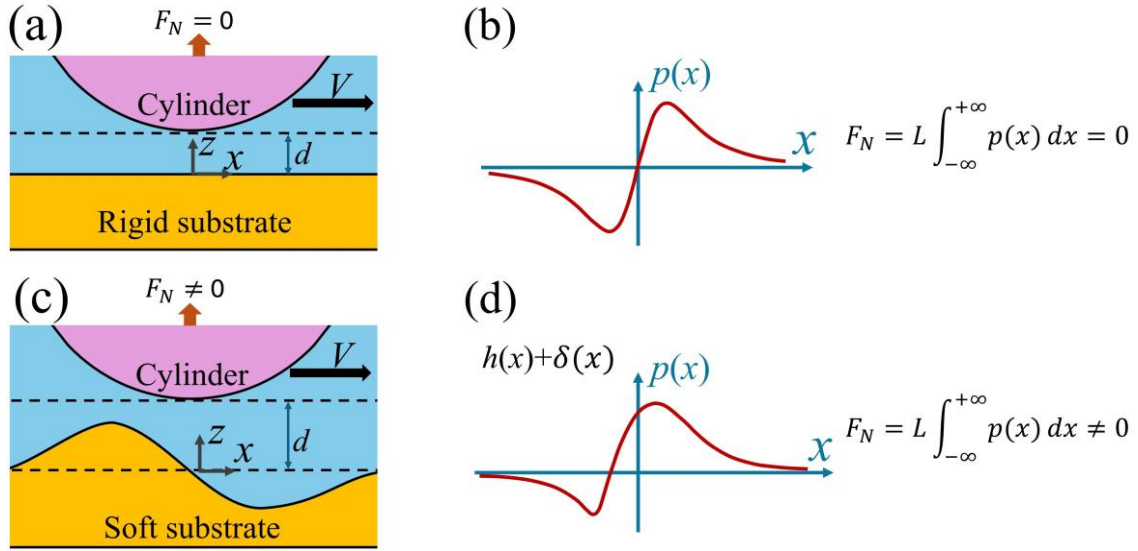

***Figure 1***: *(a) A rigid cylinder sliding laterally along a rigid substrate. (b) Pressure distribution for a cylinder moving along the rigid substrate, resulting in a normal force per unit cylinder length equal to zero. (c) A cylinder sliding laterally across a deformable substrate, causing deformation of the substrate. (d) Pressure profile for a cylinder moving along a deformable substrate, resulting in a non-zero normal force per unit length.*

Theoretical investigation of the EHD effect has been conducted over decades. From the 1990s, researchers began exploring EHD phenomena in soft matter systems (e.g. polymers, gels, biological tissues) [2-4]. A landmark contribution came from *Ken Sekimoto* and *Ludwik Leibler* in 1993 [5], who considered polymer-coated surfaces under shear. They theoretically showed that EHD coupling can cause a normal lift force and effective thickening of polymer lubricant films. *J. M. Skotheim* and *L. Mahadevan* [1, 6] derived the expression for the lift force acting on a rigid cylinder sliding along an elastic substrate. *J. Urzay* et al. [7] extended the study from the cylinder object to spherical object. *J. Beaucourt* et al. [8] reported that deformable vesicles near walls exhibit optimal lift at moderate wall compliance, the estimated vesicles modulus lies within the physiological range. *Arash Kargar* and *Bhargav Rallabandi* provided a general framework for the case of viscoelastic substrates [9] and soft patterned surfaces [10].

An important extension of EHD effect is when the "soft boundary" is not a solid but a fluid interface. A fluid–fluid interface (such as a liquid–air or liquid–liquid boundary) can be considered as a deformable membrane governed by the surface tension rather than the shear modulus $G$. Just as with a soft wall, the deformation of a fluid interface breaks the front–back symmetry of Stokes flow and can produce a net lift force on the particle. Early theoretical work by *Berdan & Leal* [11] treated the motion of a sphere near a deformable interface, finding that even a slight interfacial deformation alters the drag and can induce transverse force components. *Yang* and *Leal* [12] later examined a drop moving near a deformable interface, further elucidating how interfacial mobility and deformation affect particle motion. These studies highlighted that a nearby free surface effectively behaves like a compliant boundary – one with a capillary compliance. *Jha* et al. [13] used the lubrication approximation to calculate the analytical expressions of forces and torques felt by a rigid sphere moving near a liquid-liquid interface. *Hu* et al. [14] considered a cylinder immersed in viscous fluid moving near a flat substrate covered by an incompressible viscoelastic fluid layer, and found that shear stress gives important contributions to the lift force. This situation can be investigated by considering interfaces covered with surfactants or viscoelastic films of finite thickness [15-18].

Experimental observations of EHD lift force have advanced rapidly in the past decade. *Chan* and *Leal* [19] conducted experiments with droplets in a Couette apparatus and observed that the drops moved away from the walls toward the center of the shear flow over time. *Takemura* et al. [20] studied bubbles rising in a viscous liquid close to a vertical wall. They measured both drag and lift on the bubbles and found that as a bubble approaches the wall, it experiences a significant lift pushing it away from the wall. *Saintyves* et al. [21] investigated the sliding of a macroscopic cylinder immersed in a fluid along an inclined plane precoated with a thin gel layer, and showed that the lift force leads to a reduction in friction. *Rallabandi* et al. [22] reported a pronounced normal drift during the sedimentation of a macroscopic object along a vertical membrane under tension, highlighting the strong amplification of this effect for highly compliant, slender boundaries. *Sawaguchi* et al. [23] investigated the effect of lift force on droplet levitating over a moving wall. *Davies* et al. [24] used particle tracking in microfluidics to observe microspheres migrating away from a soft coated wall, providing quantitative validation of EHD lift scaling at the microscale. *Fares* et al. *[25]* analyzed the Brownian motion of an oil droplet near a rigid wall, from which they extracted the non-conservational force arising from EHD coupling during thermal fluctuations. More precisely, in 2020, our group reported the first direct measurement of the lift force acting on a sphere moving within a viscous liquid, near and along a polydimethylsiloxane substrate using Atomic Force Microscope (AFM) [26].

In present work, we have used the same apparatus that was used in *ref*. [26] to perform the direct measurement of the lift force acting on particles moving close to a liquid-liquid interface. To our knowledge, the colloidal AFM method is the only robust method that offers a direct measurement of the force. This setup requires only an accurate calibration of the AFM cantilever, which makes it both effective and reliable for our experiments. While the key data were previously reported in a brief Letter [27], the present work provides a comprehensive account of the experimental methodology and offers a more in-depth analysis and discussion of the underlying physics.

# 2. Theoretical model for a sphere moving along a liquid-liquid interface

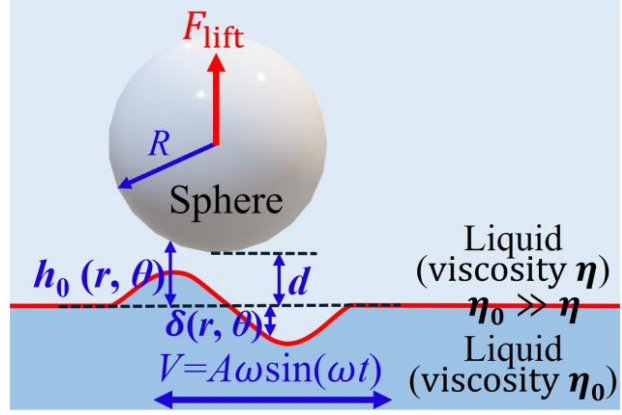

*Figure 2*. Schematic of a liquid-liquid interface formed between two immiscible liquids with different viscosities. The lower liquid moves with a velocity $V$, resulting in a lift force acting on a sphere of radius $R$.

We consider a Newtonian, incompressible fluid under laminar flow conditions (low Reynolds numbers) and assume a steady flow (neglect temporal variation of velocity). The confined fluid flow is described by the Navier-Stokes equations, which, within the framework of the lubrication approximation ($d \ll R$), can be expressed in polar coordinate as:

$$\eta \frac{\partial^2 v_r}{\partial z^2} = \frac{\partial p}{\partial r} \qquad (1)$$

$$\eta \frac{\partial^2 v_\theta}{\partial z^2} = \frac{1}{r}\frac{\partial p}{\partial \theta} \qquad (2)$$

$$\frac{\partial p}{\partial z} = 0 \qquad (3)$$

where $\eta$ is the viscosity of the fluid between the liquid interface and the sphere, $v_r$ and $v_\theta$ are radial and azimuthal components of the velocity of the fluid, $p$ is the pressure.

On the surface of the sphere and on the liquid interface, the lateral velocity of the fluid satisfies the non-slip boundary conditions:

$$v_r(z=h) = 0, \quad v_\theta(z=h) = 0 \qquad (4)$$

$$v_r(z=0) = V\cos\theta, \quad v_\theta(z=0) = -V\sin\theta \qquad (5)$$

The gap thickness is given by:

$$h(r,\theta) = d + \frac{r^2}{2R} + \delta(r,\theta) \qquad (6)$$

Integrating **Eq.** (**1**) and using the boundary conditions **Eq.** (**4**) and **Eq.** (**5**) we obtain the expression of $v_r(r,\theta,z)$:

$$v_r = \frac{1}{2\eta}\frac{\partial p}{\partial r}(z-h)z + V\cos\theta\left(1-\frac{z}{h}\right) \qquad (7)$$

Integrating **Eq.** (**2**) and using **Eq.** (**4**) and **Eq.** (**5**) we get:

$$v_\theta = \frac{1}{2\eta}\frac{1}{r}\frac{\partial p}{\partial \theta}(z-h)z + V\sin\theta\left(\frac{z}{h}-1\right) \qquad (8)$$

The continuity equation can be written in cylindrical coordinates as:

$$\frac{1}{r}\frac{\partial(rv_r)}{\partial r} + \frac{1}{r}\frac{\partial v_\theta}{\partial \theta} + \frac{\partial v_z}{\partial z} = 0 \qquad (9)$$

Integrating **Eq**. (**9**) we get

$$v(z=h) - v(z=0) = -\int_0^h \left(\frac{1}{r}\frac{\partial(rv_r)}{\partial r} + \frac{1}{r}\frac{\partial v_\theta}{\partial \theta}\right)dz \qquad 10$$

Assuming that the velocity of the variation of the gap is very small we get:

$$\int_0^h \left(\frac{1}{r}\frac{\partial(rv_r)}{\partial r} + \frac{1}{r}\frac{\partial v_\theta}{\partial \theta}\right)dz = 0 \qquad 11$$

Inserting **Eq**. (**7**) and **Eq**. (**8**) in **Eq**. (**11**) and integrating, we get the lubrication equation [7]:

$$\frac{1}{r}\frac{\partial}{\partial r}\left(rh^3\frac{\partial p}{\partial r} - 6\eta rhV\cos\theta\right) + \frac{1}{r^2}\frac{\partial}{\partial \theta}\left(h^3\frac{\partial p}{\partial \theta} + 6\eta rhV\sin\theta\right) = 0 \qquad 12$$

## 2.1 Case of rigid interface

The expression of the hydrodynamic pressure can be calculated by assuming in **Eq**. (**12**), $\delta(r,\theta) = 0$ and $h_0(r) = d + \frac{r^2}{2R}$, and thus [28]:

$$\frac{1}{r}\frac{\partial}{\partial r}\left(rh_0^3\frac{\partial p_0}{\partial r} - 6r\eta h_0 V\cos\theta\right) + \frac{1}{r^2}\frac{\partial}{\partial \theta}\left(h_0^3\frac{\partial p_0}{\partial \theta} + 6r\eta h_0 V\sin\theta\right) = 0 \qquad 13$$

By integrating **Eq**. (**13**) and using the boundary condition, $\lim_{r\to\infty} p_0(r) = 0$ we obtain the hydrodynamic pressure for a sphere moving parallel to a rigid wall [29]:

$$p_0(r,\theta) = \frac{-6\eta Vr\cos\theta}{5h_0^2} \qquad 14$$

## 2.2 Case of liquid-liquid interface

**Figure 2.** presents a schematic illustration of the liquid-liquid interface configuration. The hydrodynamic pressure between the sphere and the liquid interface is given by **Eq**. (**12**) and the deformation profile is described by the Young-Laplace equation:

$$p(r,\theta) = \sigma\nabla^2\delta = \sigma\frac{1}{r}\frac{\partial}{\partial r}\left(r\frac{\partial\delta}{\partial r}\right) + \frac{1}{r^2}\frac{\partial^2\delta}{\partial\theta^2} \qquad 15$$

where $\sigma$ is the surface tension of the interface.

To solve the set of equation **Eqs**. (**12**) and (**15**) we use the expansion of the pressure:

$$p(r,\theta) = p_0(r,\theta) + \overbrace{p_1(r,\theta)}^{\varepsilon} + \ldots \qquad 16$$

$$h(r,\theta) = h_0(r) + \delta(r,\theta) = h_0(r) + \overbrace{\delta_1(r,\theta)}^{\varepsilon} + \ldots \qquad 17$$

Injecting **Eq**. (**16**) and **Eq**. (**17**) in **Eq**. (**12**) and using **Eq**. (**13**) we obtain the following expression at first order in the pressure:

$$\frac{1}{r}\frac{\partial}{\partial r}\left(rh_0^3\frac{\partial p_1}{\partial r}\right) + \frac{h_0^3}{r^2}\frac{\partial^2 p_1}{\partial \theta^2} = \frac{6\eta V\cos\theta}{r}\frac{\partial(r\delta_1\cos\theta)}{\partial r} - \frac{3}{r}\frac{\partial}{\partial r}\left(rh_0^2\delta_1\frac{\partial p_0}{\partial r}\right) - \frac{3h_0^2}{r^2}\frac{\partial^2(\delta_1 p_0)}{\partial\theta^2} - \frac{6\eta V}{r}\frac{\partial(r\delta_1\cos\theta)}{\partial\theta}$$

$$18$$

To solve this equation, we need the expression of the interface deformation $\delta_1(r,\theta)$ that we calculate from the Young-Laplace equation. At leading order, **Eq. (15)** becomes:

$$p_0(r,\theta) = \frac{\sigma}{r}\frac{\partial}{\partial r}\left(r\frac{\partial}{\partial r}\delta_1(r,\theta)\right) + \frac{\sigma}{r^2}\frac{\partial^2}{\partial \theta^2}\delta_1(r,\theta) \qquad 19$$

We seek the solution for $\delta_1$ in the form $\delta_1(r,\theta) = \delta_1(r)\cos\theta$

and we get:

$$\frac{1}{r}\frac{\partial}{\partial r}(r\delta_1) = \frac{6\eta VR}{5\sigma\left(d+\frac{r^2}{2R}\right)} \qquad 20$$

and, thus we obtain

$$\delta_1(r) = \frac{6\eta VR^2}{5\sigma}\left(\frac{\ln\left(1+\frac{r^2}{2Rd}\right)}{r}\right) \qquad 21$$

with

$$\delta_1(r,\theta) = \delta_1(r)\cos\theta \qquad 22$$

Once we have the expression of the deformation of the interface, we can calculate the pressure to first order and the force exerted on the sphere.

We seek a solution for the pressure of the form:

$$p_1(r,\theta) = p_1^0(r) + p_1^1(r)\cos\theta + p_1^2(r)\cos2\theta + \cdots \qquad 23$$

Since the force acting on the sphere is given by:

$$F = \int_0^{+\infty}\int_0^{2\pi} rp_1(r,\theta)\,d\theta\,dr \qquad 24$$

Only the isotropic term $p_1^0(r)$ contributes to the force, while the other term that depends on the angle $\theta$ gives a zero contribution ($\int_0^{2\pi}\cos\theta\,d\theta = 0$, $\int_0^{2\pi}\cos2\theta\,d\theta = 0$).

Injecting **Eq. (23)**, **Eq. (21)** and **Eq. (22)** in **Eq. (18)** and using $(\cos\theta)^2 = \frac{1}{2} + \frac{1}{2}\cos2\theta$ we get the differential equation for the isotropic term of the pressure at first order.

We use the notation:

$$p_0(r,\theta) = p_0(r)\cos\theta$$
$$\delta_1(r,\theta) = \delta_1(r)\cos\theta$$

where, $h_0(r) = d + \frac{r^2}{2R}$, $\delta_1(r) = \frac{6\eta VR^2}{5\sigma}\left(\frac{\ln\left(1+\frac{r^2}{2Rd}\right)}{r} - \frac{r}{2Rd}\right)$ and $p_0(r) = \frac{-6\eta Vr}{5h_0^2}$.

Then we get for the isotropic term:

$$r^2\frac{\partial p_1^0(r)}{\partial r} = \frac{3\eta Vr^2\,\delta_1(r)}{h_0^3(r)} - \frac{3r^2\,\delta_1(r)}{2\,h_0(r)}\frac{\partial p_0(r)}{\partial r} \qquad 25$$

Using integration by parts, we can show that the lift force acting on the sphere is given by:

$$F_{lift} = \int_0^{+\infty} 2\pi r p_1^0(r)\,dr = -\pi\int_0^{+\infty} r^2 \frac{\partial p_1^0(r)}{\partial r}\,dr \qquad 26$$

Injecting **Eq. (25)** in **Eq. (26)** we get:

$$F_{lift} = \frac{6\pi}{25}\frac{\eta^2 V^2 R^3}{\sigma d^2} \qquad 27$$

In our experiment to produce the lateral velocity, we vibrate the substrate with an amplitude $A$ and at radial frequency ω. Since the lift force depends as the square of the velocity, we can express the force as the sum of time independent term (time average component) and a time oscillating term with a radial frequency 2ω, ( $(A\omega \cos\omega t)^2 = \frac{(A\omega)^2}{2} + \frac{(A\omega)^2 \cos 2\omega t}{2}$ ).

The lift force for oscillating velocity is then given by:

$$F_{lift} = \frac{3\pi}{25} \frac{\eta^2 (A\omega)^2 R^3}{\sigma d^2} (1 + \cos 2\omega t) \qquad 28$$

In our measurement, we focus on the first term (time average component) that we obtain by measuring the time average force applied on the sphere. Thus, **Eq**. (27) becomes:

$$F_{lift} = \frac{3\pi}{25} \frac{\eta^2 (A\omega)^2 R^3}{\sigma d^2} \qquad 29$$

# 3. Experiment

## 3.1. Experimental setup and method

**Figure 3a** presents a schematic diagram of the experimental setup. To measure the hydro-capillary lift force near a liquid-liquid interface, we employed an AFM cantilever to which a rigid borosilicate sphere was affixed. The relative lateral velocity between the interface and the sphere was obtained using a piezoelectric stage. The experiments were carried out using a Bruker Bioscope AFM equipped with a liquid-environment cantilever holder (DTFML-DD-HE). All the experiments and measurements are performed at the controlled temperature of 20 °C.

The liquid-liquid interface was formed by first placing a droplet of glycerol (≥99.5%, Sigma-Aldrich) onto a mica substrate. Subsequently, a second droplet of silicone oil was carefully layered atop the glycerol droplet, as depicted in **Figure 3b**. Initially, a 10 µL droplet of glycerol was deposited and allowed to spread, forming a "pancake" approximately 10 mm in diameter. Subsequently, a 200 µL droplet of silicone oil was deposited on top of the glycerol pancake. This arrangement created the interface necessary for measuring the lift force. The experiment was performed at the center of the interface. Due to the large size of the glycerol pancake, minor misalignments during setup did not significantly impact the measurements. In our experiment, an XY manual translation stage with a spatial resolution of a few micrometers was used to adjust the relative position between the cantilever and the pancake. The alignment was continuously monitored using the AFM optical system to ensure that the sphere was positioned above the central region of the pancake.

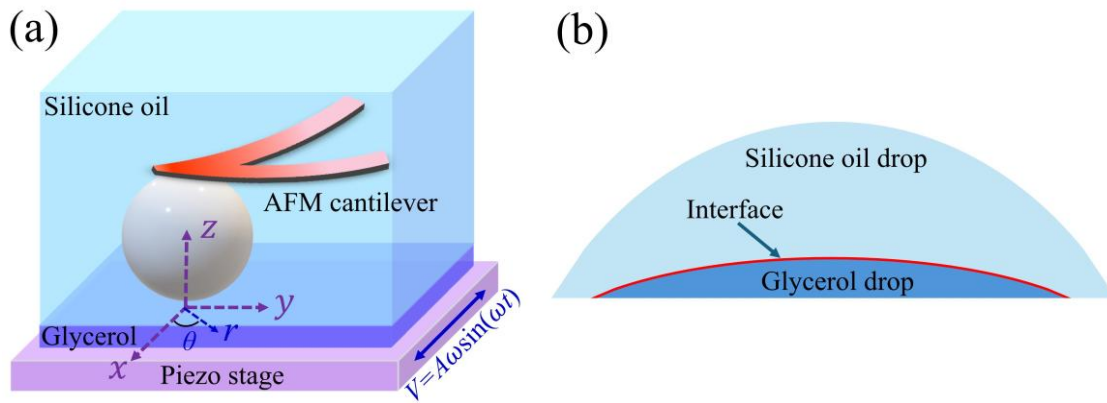

*Figure 3. (a) Diagram of the experimental setup for measuring the lift force arising from the interaction between a sliding rigid sphere and a liquid-liquid interface using an AFM-based system. A rigid borosilicate sphere is attached to the tip of an AFM cantilever. The soft interface is formed between the silicone oil (upper layer) and glycerol (lower layer), which are placed on a solid substrate fixed to a piezo stage capable of inducing lateral oscillations. The motion generates flow that deforms the interface, and the lift force acting on the sphere is measured via the vertical deflection of the AFM cantilever. (b) The liquid-liquid interface was established by layering a large drop of silicone oil over a small glycerol drop.*

Two borosilicate spheres (MO-Sci Corporation) with radii of $R = 53 \pm 1$ μm and $36 \pm 1$ μm were used as shown in **Figure 4**. Before attachment, the spheres were meticulously cleaned using ethanol and deionized water. They were then attached to the tip of a triangular silicon nitride cantilever (Model: SNL-10, Bruker) using an epoxy adhesive (Araldite, Bostik, Coubert).

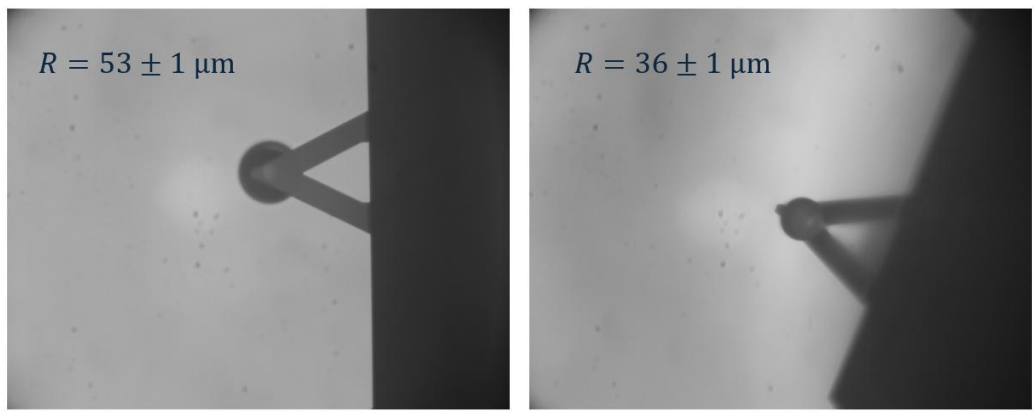

*Figure 4. Optical images of the sphere which was glued to the end of a cantilever.*

Two types of silicone oils (Silicone oil AR 20 and Silicone oil AS 100, Aldrich) with viscosities of 20 mPa·s and 100 mPa·s were used in the experiments. The liquid-liquid interface was fixed on a mica substrate mounted onto a multi-axis piezo system (NanoT series, Mad City Labs). This system was utilized to generate shear flow by inducing lateral motion of the sample perpendicular to the cantilever beam. The setup enabled the achievement of high lateral velocities by oscillating the substrate laterally. The lateral velocity amplitude $A\omega$ was controlled by adjusting either the radial frequency or the amplitude of the substrate's oscillation. In our experiments, two working frequencies, 5 Hz and 10 Hz, were used, with the amplitudes $A$ adjusted to produce lateral amplitude velocities ranging between 1.22 and 3.06 mm/s.

The multi-axis piezo system also enables precise adjustment of the gap distance $d$ between the sphere and the sample by vertically moving the substrate. To reduce the influence of drainage forces induced by piezo vertical displacement and the potential cantilever drift, vertical displacements are applied using a slow ramp motion, with amplitudes varying between 5 and 20 μm and a frequency of 5 mHz.

## 3.2. Calibration of the piezo

In this work, a piezo from MAD CITY LABS with an extensive displacement range was employed to regulate the sample's motion in both horizontal and vertical orientations. It is essential to calibrate precisely the piezo displacements in relation to the applied voltage to acquire accurate quantitative data.

### 3.2.1. Calibration for vertical motion

For the vertical movement calibration, we used an optical imaging setup to track the displacement of a slide with a calibrated grid. We applied voltages to drive the piezo's movement at a low frequency (less than 1Hz). A slide containing calibrated grids was installed in the piezo, and we recorded the motion of the slide in the vertical direction. Firstly, we calibrated the pixels by imaging the calibrated grids without any voltage applied. Then we apply a voltage to the piezo and the amplitude of the displacement can be obtained from the image as shown in **Figure 5**. We applied various voltages and measured the corresponding amplitudes of the piezo. The results are shown in **Figure 6**. By linear fitting of the results, we determined that the piezo displacement per volt was 4.9 ± 0.03 μm/V.

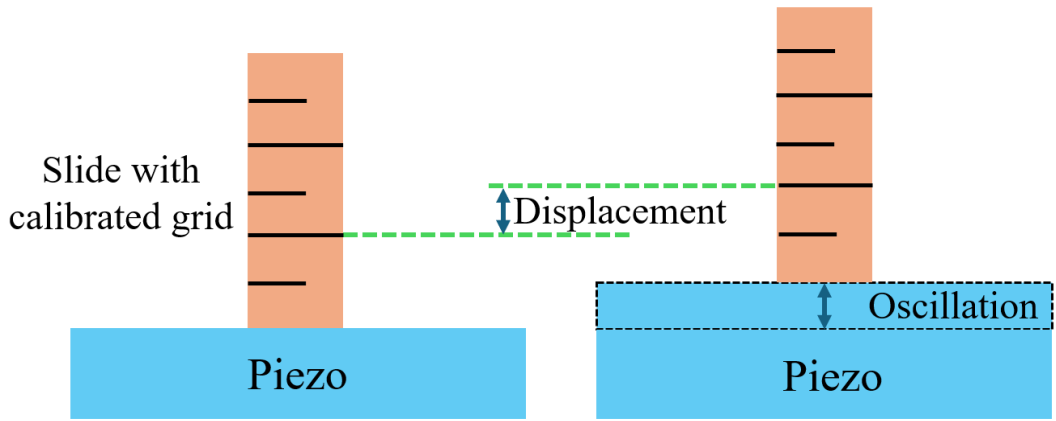

*Figure 5*. A slide with calibrated grid was fixed on the piezo, and a voltage was applied to induce the piezo displacement vertically. By tracking the slide, the displacement of the cantilever in vertical direction can be obtained for each value of the applied voltage.

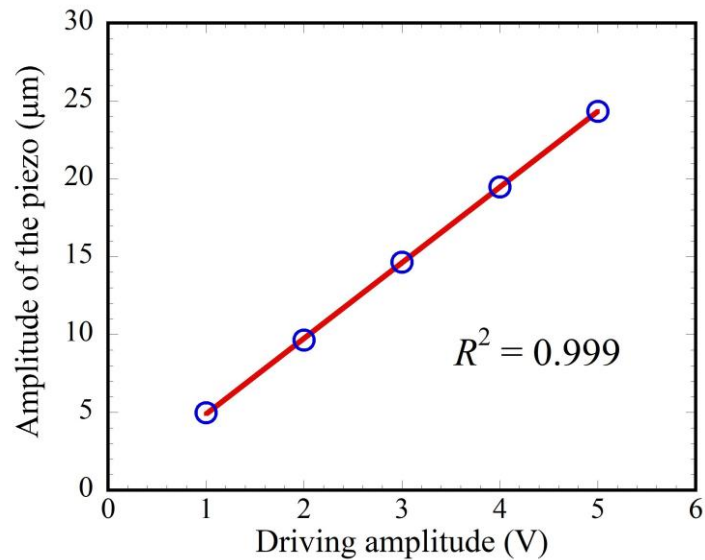

*Figure 6*. The measured amplitudes of the piezo displacement in response to driving amplitudes in volts (at 0.1 Hz). The red line represents the linear fitting curve, from which we determined the piezo displacement $4.9 \pm 0.03\ \mu m/V$.

### 3.2.2. Calibration for lateral oscillation

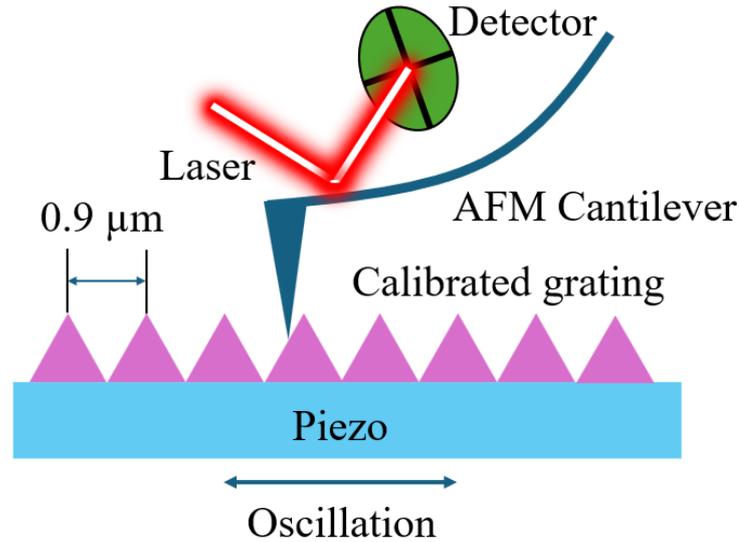

*Figure 7. The schematic of the calibration of the piezo using the grating.*

The lateral velocity between the sample and the cantilever tip is induced by a lateral displacement of the sample. Accurate calibration of this lateral movement is necessary to achieve precise knowledge of the value of the lateral velocity.

We employed a calibrated grating to evaluate the movement of the piezo. The grating has a triangular topographic profile with a periodic length of 0.9 μm, and it was attached to the piezo as shown in **Figure 7**. A cantilever with a sharp tip was brought to make contact with the grating during lateral oscillation of the piezo and then both the cantilever deflection and the piezo driving voltage were recorded simultaneously.

**Figure 8** illustrates an example of the cantilever deflection (red) and the recorded driving voltage (blue) as a function of time at a frequency of 0.5 Hz and drive amplitude of 0.5 V. By counting the peaks number in the cantilever deflection signal, the piezo displacement can be calculated by multiplying the number of peaks by the periodic length. From **Figure 8** we count 42.6 peaks corresponding to the displacement of $0.9\ \mu m \times 42.6 = 38.34\ \mu m$. The driving amplitude of the piezo is 0.49 V, thus we obtain a calibration of the lateral displacements of $38.34\ \mu m/(0.49\ V \times 2 \times 2) = 19.56\ \mu m/V$.

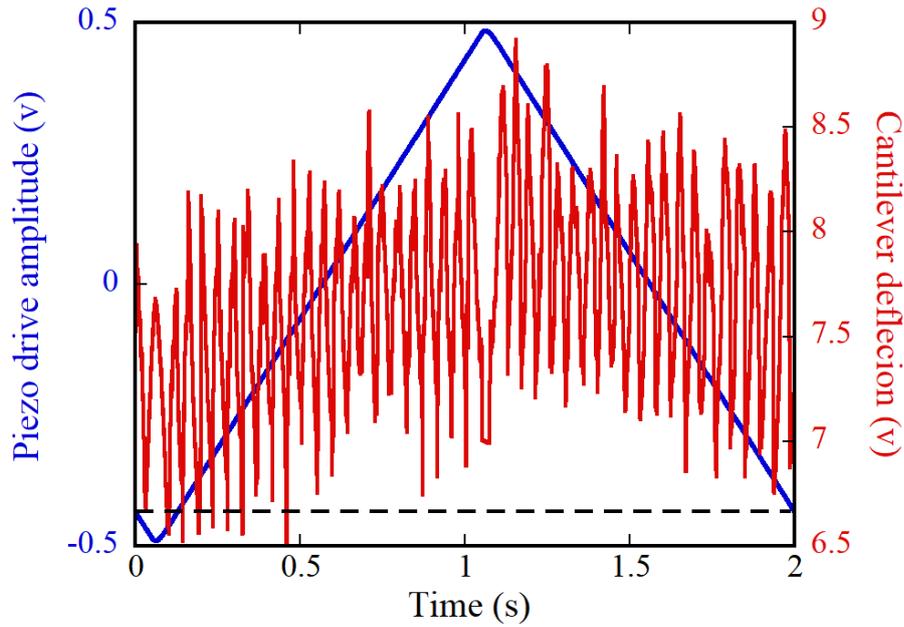

*Figure 8. The recorded cantilever deflection (red) and piezo driving voltage (blue) versus time.*

### 3.3. Calibration the cantilever stiffness

The stiffness of the colloidal probes was determined using the drainage method outlined in reference [30]. The drainage method relies on the hydrodynamic interaction between the AFM probe and the solid surface. In liquid medium, a microsphere approaching a flat surface with a velocity of $v$ induces a squeezing out of the liquid. This process generates a force which is called the hydrodynamic drainage force. The drainage method can be introduced to calibrate the cantilever by the measurement of the hydrodynamic drainage force acting on the cantilever performing vertical displacement relative to solid surface. **Figure 9** illustrates a sphere of radius $R$ moving towards a hydrophilic surface in a viscous liquid at a velocity of $V$.

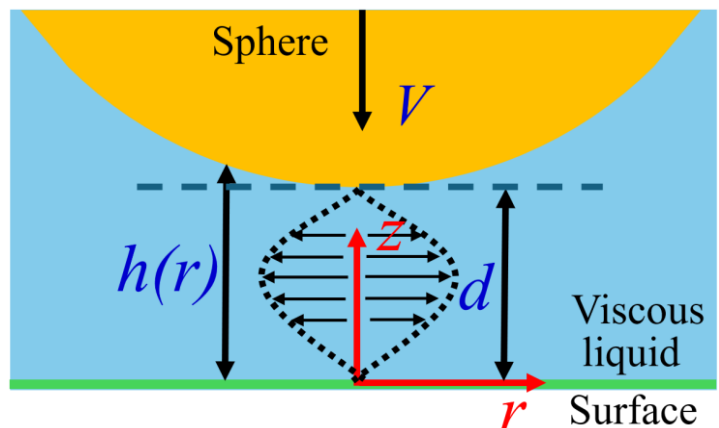

*Figure 9. Schematic illustration of cantilever calibration using the drainage force method. A spherical colloidal probe, with a radius R, approaches a hydrophilic surface at a velocity V in the z direction within a viscous liquid.*

The drainage hydrodynamic force acting on the sphere is given by:

$$F_h = -\frac{6\pi\eta R^2}{d}V \qquad 30$$

where $\eta$ the viscosity, $R$ the radius of the sphere, $d$ the distance, and $V$ the velocity of approach.

**Figure 10**a and **b** present an example of the measured deflection and the relative velocity versus the separation distance, respectively. The measurements show that the force increases with decreasing the separation distance. The relative velocity is given by the time derivative of the gap distance, and it decreases as the separation diminishes. Indeed, the relative velocity is given by the applied approach velocity minus the velocity of the cantilever deflection. At large distance the deflection is small, and the relative velocity is equal to the applied velocity of approach.

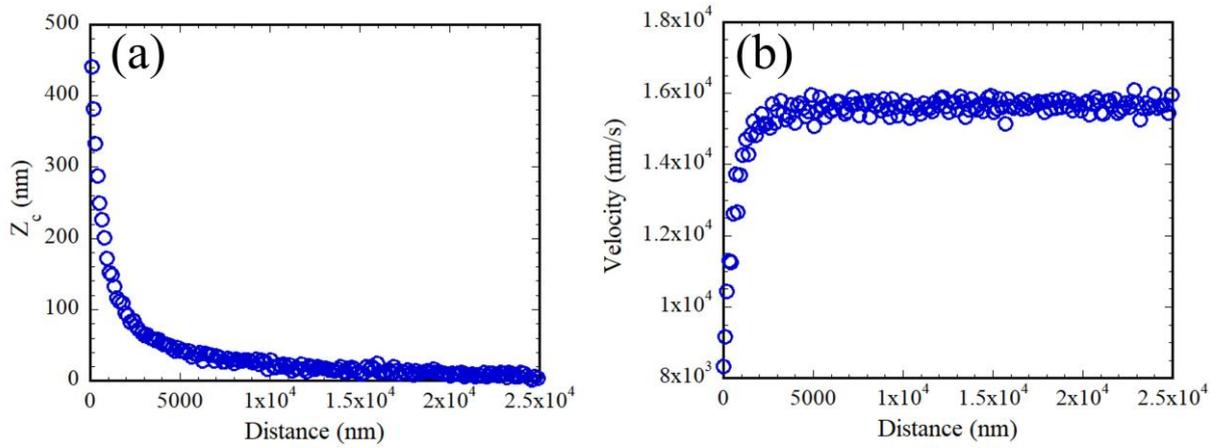

***Figure 10**. The cantilever deflection (a) and relative velocity (b) versus distance. The relative velocity $V$ is obtained from the time derivative of the separation distance.*

Equating the measured force $F = k_c Z_c$ to the hydrodynamic force, we get

$$k_c Z_c = \frac{6\pi\eta R^2}{d}V \Rightarrow \frac{V}{Z_c} = \frac{k_c}{6\pi\eta R^2}d \qquad 31$$

The spring constant $k_c$ can be obtained by linearly fitting $\frac{V}{Z_c}$ versus distance $d$, given the known of the values of viscosity and sphere radius.

In this experiment, two cantilevers (SNL-10, Bruker) with different sphere sizes were used. Calibration was performed by approaching a mica surface in viscous silicone oil with a viscosity of 100 mPa·s for the smaller sphere (36 $\mu m$ radius) and 20 mPa·s for the larger sphere (53 $\mu m$ radius). The drainage experiments were performed using the cantilever approaching a hard mica surface at a velocity of 15 $\mu m/s$, The measured $\frac{V}{Z_c}$ versus $d$ for the two cantilevers is shown in **Figure 11**. The cantilever stiffness values with an attached sphere were determined to be $k_c = 0.16 \pm 0.02$ N/m and $k_c = 0.20 \pm 0.02$ N/m for spheres with radii of $R = 36$ μm and $R = 53$ μm, respectively.

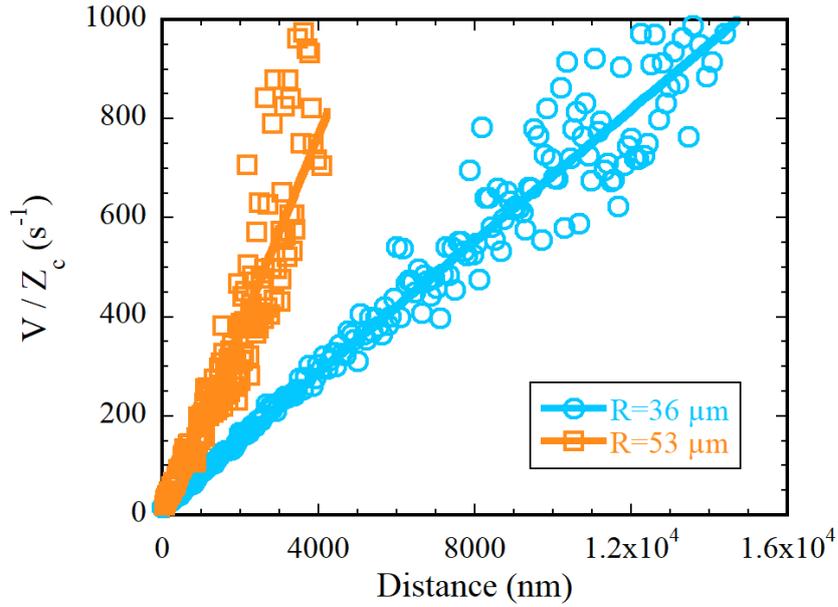

***Figure 11****. The velocity is divided by deflection versus distance. The solid lines are the linear fitting by **Eq.** (31). We obtain the stiffness $k_c = 0.16 \pm 0.02$ N/m and $k_c = 0.20 \pm 0.02$ N/m for spheres with radii of $R = 36$ μm and $R = 53$ μm, respectively.*

### 3.4. Measurement of the interfacial tension

The pendant drop technique is frequently employed for determining the interfacial tension of liquids. The formation of large droplets or bubbles is influenced by the interplay between gravity, which causes hanging droplets to elongate, and the cohesive forces among liquid molecules that promote the creation of compact, spherical droplets. Examining the form of the droplet should enable the extraction of its surface tension. In this work, the interfacial tension was measured using the pendant drop method and analyzed using the *ImageJ* plugin [31].

In this method, a glycerol droplet was suspended in silicone oil on a needle with a radius of 0.88 mm. **Figure 12** shows experimental images illustrating the measurement of interfacial tension for the silicone oil 20-glycerol interface (a) and the silicone oil 100-glycerol interface (b). Each image includes two overlays: a blue line representing the detected droplet border, and a red line depicting the fitted profile based on the pendant drop method (note that the blue line is mostly obscured by the red line due to their close overlap). The measurements revealed that the interfacial tension was consistent, with a value of $\sigma = 21 \pm 2$ mN/m, for both the silicone oil 20-glycerol interface and the silicone oil 100-glycerol interface.

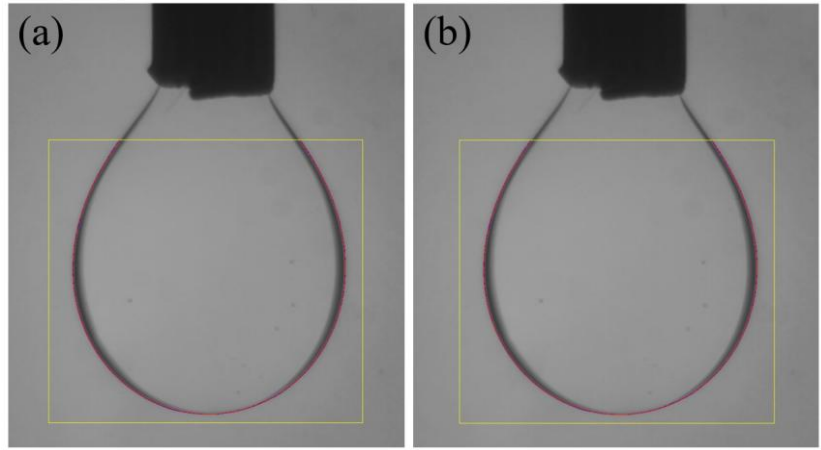

*Figure 12. Examples of pendant drop images analyzed using the ImageJ plugin for the glycerol-silicone 20 interface (a) and glycerol-silicone 100 interface (b). In each figure, the blue line indicates the detected drop border by the plugin, while the red line represents the fitting profile.*

# 4. Results and discussion

## 4.1. Effect of the interface softness

In this study, we focused on the time-independent component of the lift force, as described by **Eq**. (**29**), by analyzing the time-averaged force acting on the sphere. **Figure 13** displays the results for a sphere immersed in silicone oil 20, sliding along the interface formed with glycerol. The data is plotted as a function of the distance between the sphere and the interface. For comparison, additional measurements were conducted under identical experimental conditions, with the sphere sliding along a rigid mica substrate in silicone oil 20.

No repelling force was observed on the rigid mica substrate, as its high rigidity prevents significant deformation. At very small distances, below approximately 20 nm, a weak force attributable to drainage hydrodynamic effects was detected. As mentioned above, in our experiments the gap distance was controlled by imposing a vertical displacement of the sample at very low velocities. Even at a vertical velocity as low as 100 nm s⁻¹, the hydrodynamic drainage force (Eq. (30)) can become significant when the gap is smaller than ~20 nm. For gap distances larger than ~20 nm, this force is negligible under our experimental conditions.

For the soft liquid-liquid interface, a lift force was prominent, even at relatively large distances, and increased significantly as the gap distance decreased. These results underscore the distinct role of interface softness in generating measurable lift forces and highlight the dynamic interplay between hydrodynamic forces and interface deformation.

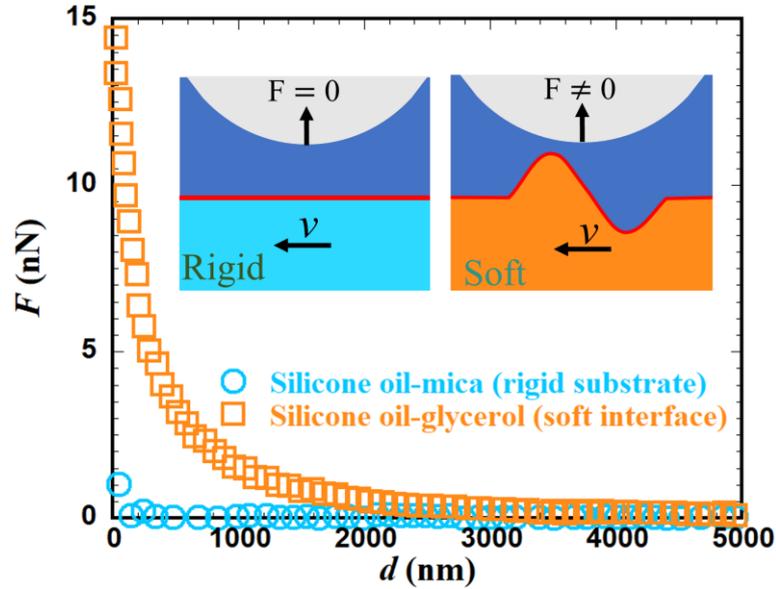

***Figure 13***. *The experiment measured the forces acting on two distinct interfaces: a soft silicone oil-glycerol interface and a rigid mica substrate. For the soft interface, the data points are represented as squares, while for the rigid substrate, circles are used. Both experiments utilized a 36 μm radius sphere in silicone oil with a viscosity of 20 mPa·s. The tests were conducted under identical conditions, with a lateral velocity amplitude of 2.59 mm/s and a working frequency of 10 Hz.*

### 4.2. Effect of velocity

**Figure 14a** illustrates the force measured as a function of the gap distance to the silicone oil-glycerol interface under varying amplitude velocities. The data reveals a clear increasing trend in force with higher amplitude velocities. **Figure 14b** presents the normalized forces versus distance in a logarithmic scale, where the force is divided by the square of the velocity. Consistent with the theoretical expression for lift force, the normalized forces collapse with each other at larger distances, indicating uniform behavior. A dashed line in the inset highlights a power law of $d^{-2}$, as expected. For these measurements, the viscosity of silicone oil is 20 mPa·s, the sphere radius is 36 μm, and the working frequency is 5 Hz. These results confirm that velocity has a significant impact on the lift force.

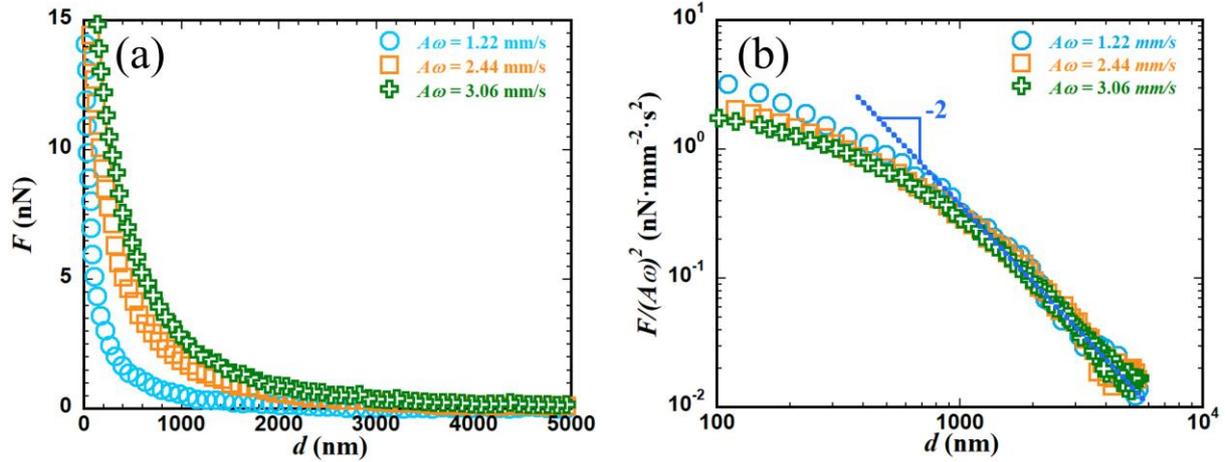

***Figure 14.*** *(a) Experiments were conducted using three different amplitude velocities while maintaining the same frequency. (b)The logarithmic scale of the normalized force versus distance, obtained by dividing the measured force by the square of the velocity. The experiments were performed with silicone oil of viscosity 20 mPa·s, a sphere radius of 36 μm, and a working frequency of 5 Hz.*

### 4.3. Effect of viscosity

**Figure 15**a shows force measurements conducted with silicone oils of two viscosities: 20 and 100 mPa·s. The results show a larger lift force with higher viscosity. Importantly, the interfacial tension between silicone oil 20 and silicone oil 100 with glycerol remains constant, isolating viscosity as the influencing parameter. At large distances, the normalized force, which is calculated by dividing the force by the square of viscosity, aligns closely between both measurements as shown in **Figure 15**b. This consistency supports the prediction from **Eq**. (**29**), derived under conditions of small interface deformation at large distances. For both experiments, the sphere radius is 36 *μm*, the amplitude of the velocity is 1.22 mm/s, and the working frequency is 5 Hz.

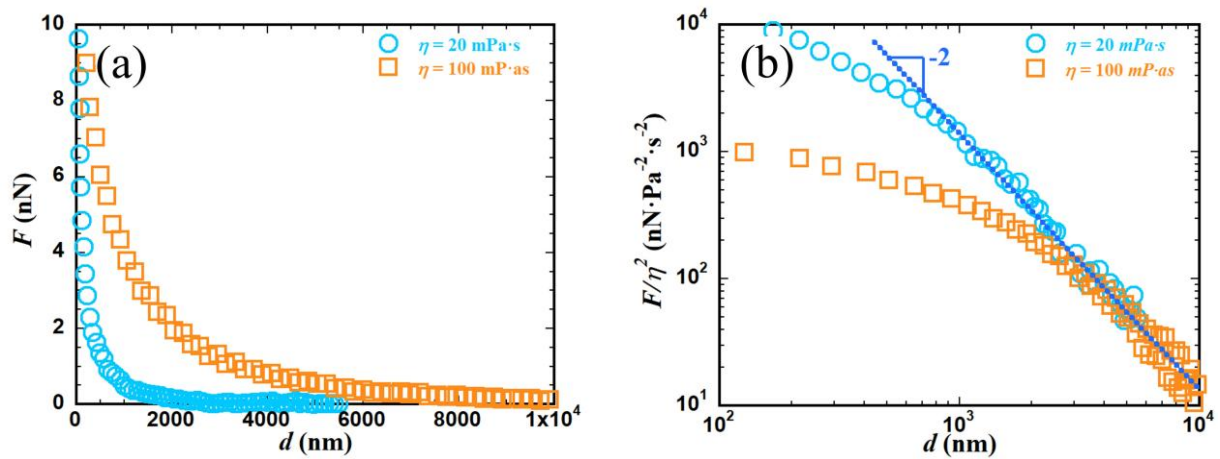

**Figure 15**. *(a) Experiments were conducted using two silicone oils with viscosities of 20 and 100 mPa·s. (b) The normalized force, calculated by dividing the force by the square of the viscosity. Both measurements were performed with a sphere of radius 36 μm, a velocity amplitude of 1.22 mm/s, and a working frequency of 5 Hz.*

### 4.4. Effect of sphere size

**Figure 16**a depicts force measurements using spheres of two different radii. As anticipated, the larger sphere results in higher lift forces due to the greater hydrodynamic interaction with the interface. **Figure 16**b illustrates the normalized force, obtained by dividing by the cube of the sphere radius, which collapses at large distances, confirming theoretical predictions. For these measurements, the viscosity of silicone oil is 20 mPa·s, the velocity amplitude is 2.59 mm/s, and the working frequency is 10 Hz for both sphere sizes. Additionally, the log-log plots in **Figure 14**b, **Figure 15**b, and **Figure 16**b reaffirm that the lift force scales with $d^{-2}$ at large distances. These measurements highlight a consistent relationship between the lift force and the cube of the radius, aligning with the expression given in **Eq**. (**29**).

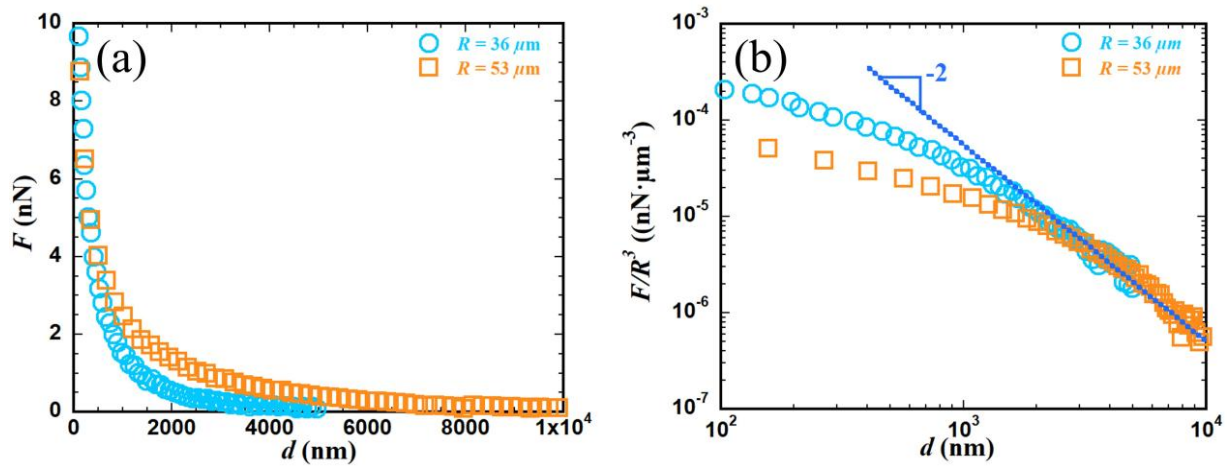

*Figure 16. (a) Experiments were performed with spheres of two different radii of 36 μm and 53 μm. (b) The inset log-log plot presents the normalized force, obtained by dividing the measured force by the cube of the radius. Both measurements were conducted using silicone oil with a viscosity of 20 mPa·s, an amplitude velocity of 2.59 mm/s, and a working frequency of 10 Hz.*

### 4.5. Lift force normalization

At small distances, the measurements do not coincide with each other. This discrepancy arises because **Eq**. (**29**), which provides the analytical expression for the lift force, is valid only at large distances. The derivation of this equation assumes small interface deformations, which is not the case at small distances. For smaller distances, the set of equations **Eq**. (**12**) and Eq. **Eq**. (**15**) must be solved numerically to accurately determine the force. To solve numerically the problem, we introduce the dimensionless variables: $\hat{p} = \frac{p}{p^*}, \hat{h} = \frac{h}{d}, \hat{r} = \frac{r}{\sqrt{Rd}}, \hat{\delta} = \frac{\delta}{d}$ where $p^* = \frac{\eta V R^{1/2}}{d^{3/2}}$ and we rewrite the set of **Eqs**. (12) and (15) in dimensionless form:

$$\frac{1}{\hat{r}}\frac{\partial}{\partial \hat{r}}\left(\hat{r}\,\hat{h}^3 \frac{\partial \hat{p}}{\partial \hat{r}} - 6\,\hat{r}\hat{h}\,\cos\theta\right) + \frac{1}{\hat{r}^2}\frac{\partial}{\partial \theta}\left(\hat{h}^3 \frac{\partial \hat{p}}{\partial \theta} + 6\hat{r}\hat{h}\,\sin\theta\right) = 0 \qquad 32$$

$$\kappa\, p(r,\theta) = \nabla^2 \delta \qquad 33$$

where $\kappa = \eta A \omega\, R^{\frac{3}{2}}/\left(\sigma d^{\frac{3}{2}}\right)$

By solving the set of **Eqs**. (12) and (15) we obtain the pressure and then the force acting on the sphere is calculated as:

$$F_{lift} = \int rp\cos(\theta)\, d\theta\, dr = F^* \int \hat{r}\hat{p}\cos(\theta)\, d\theta d\hat{r} \qquad 34$$

where $F^* = \frac{\eta V\, R^{3/2}}{d^{1/2}}$. The dimensionless parameter [6] $\kappa = \eta A \omega\, R^{\frac{3}{2}}/\left(\sigma d^{\frac{3}{2}}\right)$ characterizes the softness of the interface. At large distances, $\kappa$ is very small, indicating minimal interface deformation. Conversely, at small distances, $\kappa$ becomes large, corresponding to significant deformation of the interface.

When $\kappa \ll 1$, corresponding to the large distances, we obtain:

$$\frac{F}{F^*} = \left(\frac{3\pi}{25}\right)\kappa \qquad 35$$

which is exactly the dimensionless form of the equation **Eq**. (29).

To better understand the relationship between the lift force and the various parameters, the normalized force $F/F^*$ is plotted against the dimensionless softness parameter $\kappa$. **Figure 17** presents experimental data collected across different working frequencies, amplitude velocity, viscosities, and sphere sizes.

In the regime of small softness parameters, corresponding to larger distances, the measurements are consistent and display a linear relationship with $\kappa$, as predicted by **Eq**. (35). This behavior is represented by the continuous red curve in **Figure 17**. As $\kappa$ increases, that corresponds to smaller distances and larger deformations, the normalized force begins to saturate. While the overall trend aligns with the theoretical model, slight deviations emerge, which become more significant with higher radial frequencies, greater viscosities, and larger spheres.

At low excitation frequencies, the observed saturation can be attributed to the nonlinearity between interface deformation and velocity. This phenomenon is captured by solving the Reynolds equation (**Eq**. (32)) in conjunction with the Young-Laplace equation (**Eq**. (33)). The resulting numerical predictions, represented by the black curve in **Figure 17**, demonstrate saturation at high softness parameters and align well with the experimental data. Despite this overall agreement, minor discrepancies emerge at large values of $\kappa$.

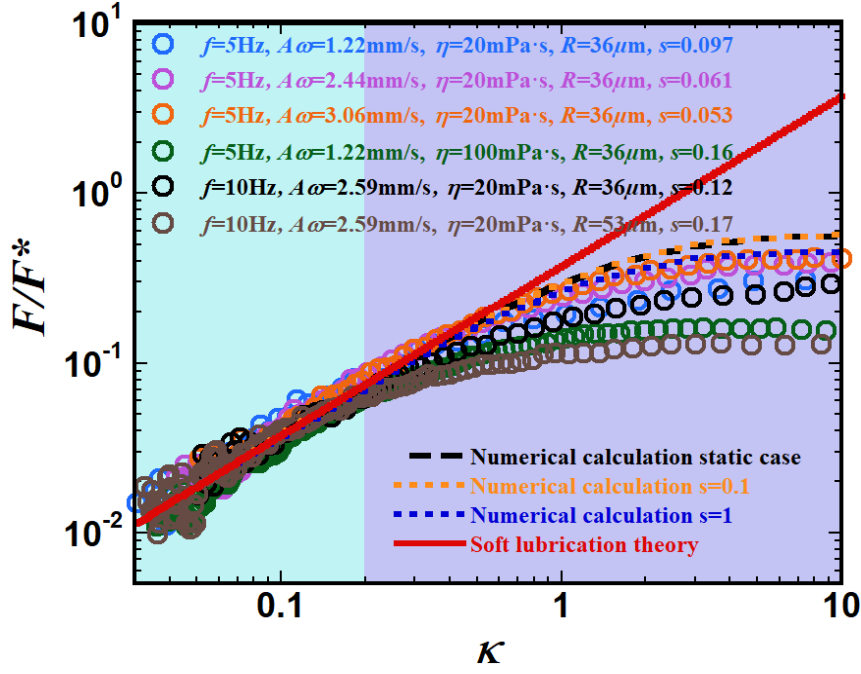

*Figure 17. The plot depicts the dimensionless force $F/F^*$ versus the dimensionless compliance $\kappa$ on a logarithmic scale. Experimental data, represented by circle markers of varying colors, correspond to measurements conducted under diverse conditions, including different amplitude velocities, working frequencies, sphere radii, and upper liquid viscosities. The continuous red curve reflects the theoretical prediction given by **Eq. (29)**, while the numerical calculations (**Eq. (36)**) for s values of 0.01 (static), 0.1, and 1 are also depicted by the black, orange, and blue curves, respectively.*

### 4.6 Effect of torque

One hypothesis for these deviations is the effect of sphere rotation caused by hydrodynamic torque. As the gap between the sphere and the interface decreases, the torque generated by the lubricated flow could induce a rotational motion of the sphere, potentially altering the relative velocity at the interface. To examine this possibility, the lateral torsion signal of the cantilever was analyzed to estimate the rotational velocity of the sphere.

To determine the velocity of the sphere induced by the torque, we first calibrate the torsion signal of the cantilever (twist). The AFM colloidal probe is brought into contact with the mica substrate immersed in silicone oil. The lateral displacement of the substrate ($S$) induces a twist ($\varphi$) of the probe, thereby introducing a torque ($\Omega$), as illustrated in **Figure 18**.

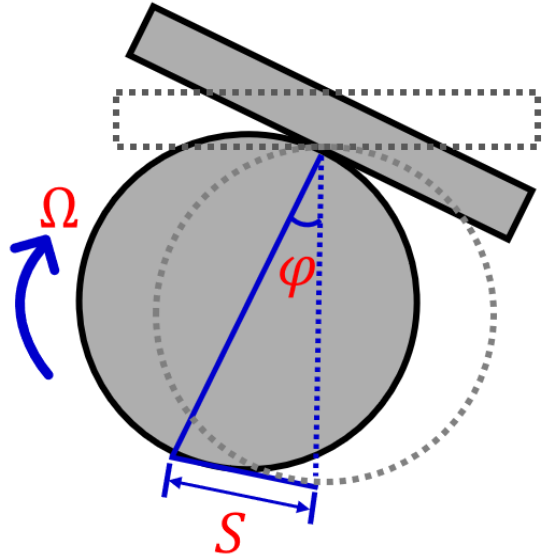

*Figure 18. The Schematic illustration for the twist of an AFM colloidal probe.*

In our experiments, a small lateral displacement of the mica substrate ($S = 61\ nm$) generated a torsion signal of 1.252 V. Using the relation $\varphi = \frac{S}{2R}$, where $R = 53\ \mu m$, the calculated $\varphi$ is $5.8 \times 10^4$. From this, we determine the sensitivity to be $5.1 \times 10^4\ rad/V$.

Consequently, in our measurements of the lift force, the hydrodynamic oscillating torsion angle ($\varphi$) can be obtained by multiplying the measured voltage signal by the sensitivity as shown in **Figure 19a**. The oscillating apex velocity of the sphere ($V_{apex}$) is then calculated using the relation $V_{apex} = 2R\frac{\partial \varphi}{\partial t}$. In **Figure 19b**, we show the ratio between the amplitude of the apex velocity and the substrate velocity. The results revealed that the rotational velocity was negligible, amounting to less than 3% of the sliding velocity. This small rotation velocity suggests that it cannot explain the deviations observed at higher softness parameters, pointing to the need for further investigation into other contributing factors.

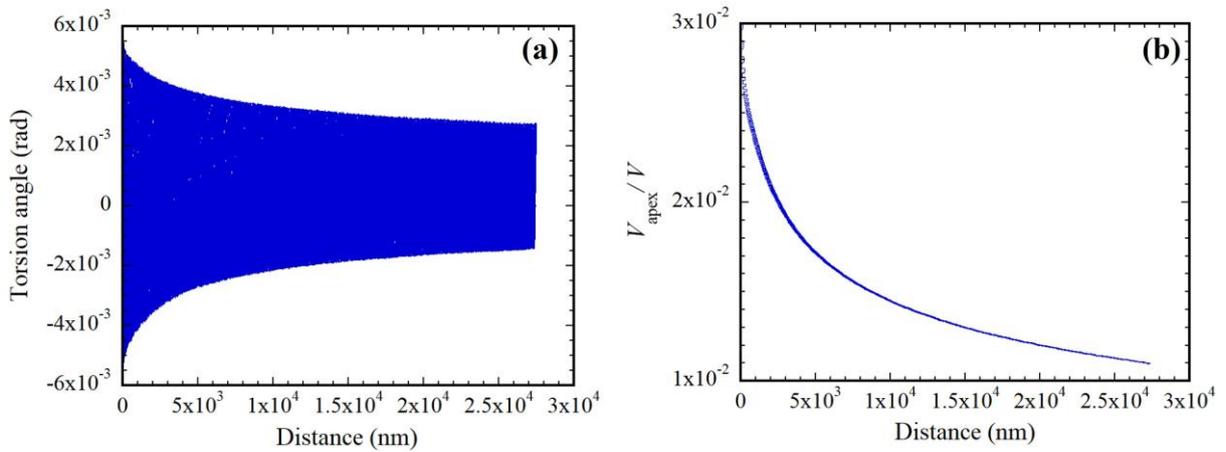

*Figure 19. (a) A typical oscillating torsion angle measurement at a substrate velocity of 1.25 mm/s. (b) The ratio of the sphere apex amplitude velocity to the substrate amplitude velocity.*

## 4.7. Non-stationary effect

For a given viscosity, at higher frequencies the experimental results diverge notably from theoretical expectations. This larger discrepancy at higher frequencies suggests that the non-stationary contribution may influence the behavior of the system at large values of $\kappa$. To examine these non-stationary effects more closely, we move away from the assumption that the velocity $v$ remains constant over time. Instead, we consider a scenario in which the interface profile evolves dynamically, imposing a boundary condition where the vertical fluid velocity at the interface is given by $-\partial_t \delta$. Incorporating this time-dependent deformation requires revisiting the dimensionless Reynolds equation (**Eq.** (32)) to account for the changing interface profile.

$$\frac{1}{\hat{r}}\frac{\partial}{\partial \hat{r}}\left(\hat{r}\ \hat{h}^3 \frac{\partial \hat{p}}{\partial \hat{r}} - 6\ \hat{r}\hat{h}\ \cos\theta\right) + \frac{1}{\hat{r}^2}\frac{\partial}{\partial \theta}\left(\hat{h}^3 \frac{\partial \hat{p}}{\partial \theta} + 6\hat{r}\hat{h}\ \sin\theta\right) = -\frac{12s}{\kappa^{\frac{1}{3}}}\frac{\partial \hat{\delta}}{\partial \tau} \qquad 36$$

where $s = (R^3\ \omega\eta/(A^2\sigma))^{\frac{1}{3}}$, and $\tau$ is the dimensionless times given as $\tau = \omega t$.

In **Figure 17**, each individual curve represents a scenario where the distance $d$ is varied while all other parameters remain constant, ensuring that each curve corresponds to a specific fixed value of $s$. Numerically solving the time-dependent partial differential equation shows that at large $\kappa$, the predicted force values reach a saturation that depends on $s$. Qualitatively, this trend is consistent with the experimental data. However, the agreement is not quantitative. In particular, theoretical predictions made with $s = 0.1$ (dashed yellow curve) produce results very similar to those obtained under static conditions (black curve).

## 4.8. Finite value of the viscosity ratio between glycerol and silicon oil

A plausible reason for the observed discrepancies between the theoretical predictions and the experimental results at large $\kappa$ is the finite value of the viscosity ratio between glycerol and silicon oil, which until now has been treated as an idealized limit. In practice, the glycerol beneath the interface possesses a finite viscosity, causing the interface to move more slowly than the velocity $v$ imposed by the motion of the mica substrate.

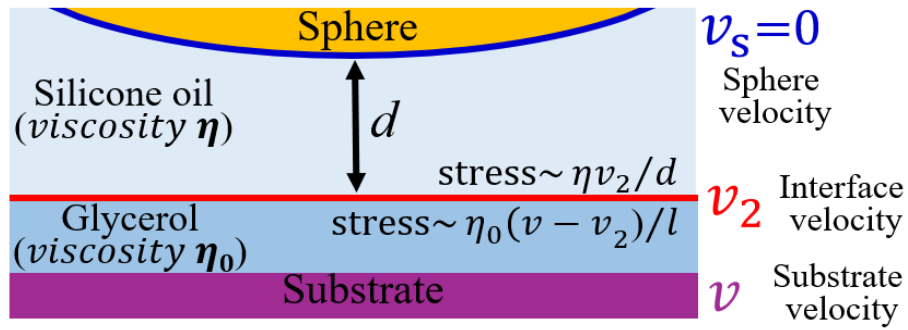

***Figure 20***. *Schematics of the velocity and shear stress at different positions, where the imposed velocity is $v$ and the interface velocity is $v_2$. The sphere is assumed to be in a static state.*

To estimate the actual interface velocity, $v_2$, one can use the fact that the lateral shear stress is equal on both sides of the interface. On the upper side, the shear stress is roughly $\sigma_{xz} \sim \eta v_2/d$. On the lower side, it takes the form $\eta_0(v - v_2)/l$, where $\eta_0$ is the viscosity of glycerol and $l$ is a characteristic length scale on the order of a few multiples of $\sqrt{2Rd}$. These two expressions lead to: $v_2 \sim \frac{v}{1+\eta l/(\eta_0 d)}$. From this relationship, it follows that the interface velocity closely approaches the imposed velocity $v$ only when $d$ is sufficiently large, specifically, when $d \gg \frac{\eta l}{\eta_0}$. In other words, the finite value of the viscosity ratio between glycerol and silicon oil significantly influences the interface dynamics, especially at smaller distances where the interface velocity can no longer be assumed equal to the imposed substrate speed. Exploring the non-stationary effects that arise when considering the finite value of the viscosity ratio between glycerol and silicon oil lies outside the scope of the study in this work. However, we believe that this effect could account for the minor deviations observed between theoretical calculations and experimental results at high softness parameters.

### 4.9. Time-average and the second harmonic component

At large sphere-interface distance, **Eq**. (**27**) is applicable because the interface deformation is small. In this regime, the force scales quadratically with the velocity. This scaling makes it straightforward to decompose the observed force into two contributions: a steady (time average component) component and a second harmonic component, both of which should theoretically match. However, as the sphere moves closer and the gap distance diminishes, physical process becomes more intricate. The interface deformation becomes large, and the force can no longer be approximated by a pure square-velocity relationship. Instead, additional harmonic components arise, extending beyond the second harmonic and complicating the overall force signature. Under these circumstances, focusing exclusively on the second harmonic term fails to capture the full complexity of the interaction.

To capture the complete physical behavior at smaller distances, we shifted our focus to the time average component of the force rather than relying exclusively on harmonic decomposition. For instance, **Figure 21** offers a direct comparison between the time-averaged force measurements and those based solely on the second harmonic. This figure highlights the contrast between the two methods. In parallel, our computational approach was adapted: rather than relying on a single harmonic representation, we incorporate time-averaged values of the calculated force. This approach allows for a more accurate comparison between experimental data and calculations, ensuring that the nuances of the non-linear, multi-harmonic force response are properly accounted for.

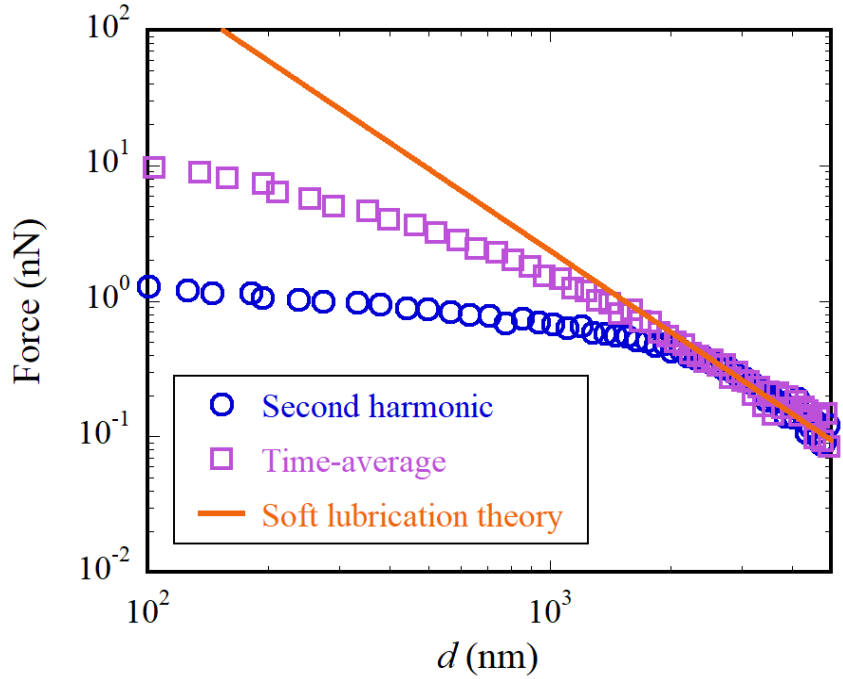

*Figure 21. The measured force obtained with the time-average (square) and with the second harmonic component (circle). The measurement was conducted at a frequency of 10 Hz and a lateral velocity amplitude (Aω) of 2.59 mm/s, using a sphere with a radius of 36 μm and silicone oil viscosity of 20 mPa·s.*

## Conclusions

This work presents direct quantitative measurements of the lift force acting on a sphere sliding along a liquid-liquid interface. We describe in detail the experimental colloidal AFM setup and its calibration, allowing high-sensitivity force measurements. We systematically investigate the parameters influencing the lift force and identify the key factors governing its behavior, and the experimental results were compared with theoretical prediction. The main findings are summarized as follows:

- No measurable lift force was detected on a sphere sliding along a rigid mica substrate. By contrast, a pronounced lift force emerged for the sphere sliding on a liquid–liquid interface as a consequence of interfacial deformation.

- For large distances, the measured lift force increases with the inverse square of the distance, the square of the velocity, the square of the viscosity, and the cube of the radius. For small distances, saturation was observed.

- By introducing the dimensionless parameter $\kappa$ (softness parameter): For low $\kappa$, all normalized force measurements collapse in the master curve and follow a linear relationship with the softness parameter, aligning with theoretical prediction. As the softness parameter increases, saturation is observed, matching numerical simulations.

- The small discrepancies observed between the experimental results and the theoretical predictions have been verified not to originate from torque effects. The discrepancies may instead arise from limitations of the theoretical model, which does not account for both the finite viscosity ratio between glycerol and silicone oil and non-stationary effects.

These experimental findings offer valuable insights into interfacial dynamics and set the stage for further exploration of object motion near liquid-liquid interfaces.

Future theoretical studies that address non-stationary effects and explicitly account for the finite viscosity ratio between the upper and lower liquids may help resolve the observed deviations at saturation, thereby providing a more comprehensive understanding of lift force behavior in such systems.

For future experimental work, we propose to investigate the lift force generated on a sphere moving along liquid-liquid interfaces covered with surfactants or viscoelastic films of finite thickness. The behavior of such a lift force is expected to differ from that reported in this paper. Indeed, for viscoelastic interfaces, the effect of tangential shear stress at the interface can play a crucial role in determining the lift force [14].

# Acknowledgments

The authors thank Aditya Jha, Yacine Amarouchene, Thomas Salez, Thomas Guerin, Chaouqi Misbah for fruitful discussion. H. Zhang acknowledge the financial support of the China Scholarship Council. The authors thank the French National Research Agency for the supporting grants MS-DOS (Grant No. ANR-25-CE06-3953-02) and EDDL (Grant No. ANR-19-CE30-0012).

# References

[1]     J.M. Skotheim, L. Mahadevan, Soft lubrication, Phys. Rev. Lett. 92(24) (2004) 245509.

[2]     I. Cantat, C. Misbah, Lift force and dynamical unbinding of adhering vesicles under shear flow, Phys. Rev. Lett. 83(4) (1999) 880.

[3]     U. Seifert, Hydrodynamic lift on bound vesicles, Phys. Rev. Lett. 83(4) (1999) 876.

[4]     M. Abkarian, C. Lartigue, A. Viallat, Tank treading and unbinding of deformable vesicles in shear flow: determination of the lift force, Phys. Rev. Lett. 88(6) (2002) 068103.

[5]     K. Sekimoto, L. Leibler, A mechanism for shear thickening of polymer-bearing surfaces: elasto-hydrodynamic coupling, Europhys. Lett. 23(2) (1993) 113.

[6]     J. Skotheim, L. Mahadevan, Soft lubrication: The elastohydrodynamics of nonconforming and conforming contacts, Phys. Fluids 17(9) (2005).

[7]     J. Urzay, S.G. Llewellyn Smith, B.J. Glover, The elastohydrodynamic force on a sphere near a soft wall, Phys. Fluids 19(10) (2007).

[8]     J. Beaucourt, T. Biben, C. Misbah, Optimal lift force on vesicles near a compressible substrate, Europhys. Lett. 67(4) (2004) 676.

[9]     A. Kargar-Estahbanati, B. Rallabandi, Lift forces on three-dimensional elastic and viscoelastic lubricated contacts, Phys. Rev. Fluids 6(3) (2021) 034003.

[10]    A. Kargar-Estahbanati, B. Rallabandi, Non-monotonic frictional behavior in the lubricated sliding of soft patterned surfaces, Soft Matter 21(3) (2025) 448-457.

[11]    C. Berdan Ii, L. Leal, Motion of a sphere in the presence of a deformable interface: I. Perturbation of the interface from flat: the effects on drag and torque, J. Colloid Interface Sci. 87(1) (1982) 62-80.

[12]    S.-M. Yang, L. Leal, Motions of a fluid drop near a deformable interface, Int. J. Multiphase Flow 16(4) (1990) 597-616.

[13]    A. Jha, Y. Amarouchene, T. Salez, Capillary lubrication of a spherical particle near a fluid interface, J. Fluid Mech. 1001 (2024) A58.

[14]    S. Hu, F. Meng, Effect of fluid viscoelasticity, shear stress, and interface tension on the lift force in lubricated contacts, J. Chem. Phys. 159(16) (2023).

[15]    L. Suswanth, G.K. Rajan, Generalized theoretical framework for spatial attenuation rates of gravity waves in a stepwise-stratified fluid system with multiple layers of arbitrary depths, Phys. Rev. Fluids 10(5) (2025) 054802.

[16]    S. Ermakov, I. Sergievskaya, L. Gushchin, Damping of gravity-capillary waves in the presence of oil slicks according to data from laboratory and numerical experiments, Izv. Atmos. Ocean. Phys. 48(5) (2012) 565-572.

[17]    I. Sergievskaya, S. Ermakov, Damping of gravity–capillary waves on water surface covered with a visco-elastic film of finite thickness, Izv. Atmos. Ocean. Phys. 53(6) (2017) 650-658.

[18]    D. Langevin, Rheology of adsorbed surfactant monolayers at fluid surfaces, Annu. Rev. Fluid Mech. 46(1) (2014) 47-65.

[19]    P.C.-H. Chan, L. Leal, An experimental study of drop migration in shear flow between concentric cylinders, Int. J. Multiphase Flow 7(1) (1981) 83-99.

[20]    F. Takemura, S. Takagi, J. Magnaudet, Y. Matsumoto, Drag and lift forces on a bubble rising near a vertical wall in a viscous liquid, J. Fluid Mech. 461 (2002) 277-300.

[21] B. Saintyves, T. Jules, T. Salez, L. Mahadevan, Self-sustained lift and low friction via soft lubrication, Proc. Natl. Acad. Sci. U. S. A. 113(21) (2016) 5847-5849.

[22] B. Rallabandi, N. Oppenheimer, M.Y. Ben Zion, H.A. Stone, Membrane-induced hydroelastic migration of a particle surfing its own wave, Nat. Phys. 14(12) (2018) 1211-1215.

[23] E. Sawaguchi, A. Matsuda, K. Hama, M. Saito, Y. Tagawa, Droplet levitation over a moving wall with a steady air film, J. Fluid Mech. 862 (2019) 261-282.

[24] H. Davies-Strickleton, D. Débarre, N. El Amri, C. Verdier, R.P. Richter, L. Bureau, Elastohydrodynamic lift at a soft wall, Phys. Rev. Lett. 120(19) (2018) 198001.

[25] N. Fares, M. Lavaud, Z. Zhang, A. Jha, Y. Amarouchene, T. Salez, Observation of Brownian elastohydrodynamic forces acting on confined soft colloids, Proc. Natl. Acad. Sci. U. S. A. 121(42) (2024) e2411956121.

[26] Z. Zhang, V. Bertin, M. Arshad, E. Raphaël, T. Salez, A. Maali, Direct measurement of the elastohydrodynamic lift force at the nanoscale, Phys. Rev. Lett. 124(5) (2020) 054502.

[27] H. Zhang, Z. Zhang, A. Jha, Y. Amarouchene, T. Salez, T. Guérin, C. Misbah, A. Maali, Direct measurement of the viscocapillary lift force near a liquid interface, Phys. Rev. Lett. 134(9) (2025) 094001.

[28] A. Prosperetti, Linear oscillations of constrained drops, bubbles, and plane liquid surfaces, Phys. Fluids 24(3) (2012).

[29] M. O'neill, K. Stewartson, On the slow motion of a sphere parallel to a nearby plane wall, J. Fluid Mech. 27(4) (1967) 705-724.

[30] V.S. Craig, C. Neto, In situ calibration of colloid probe cantilevers in force microscopy: hydrodynamic drag on a sphere approaching a wall, Langmuir 17(19) (2001) 6018-6022.

[31] A. Daerr, A. Mogne, Pendent_drop: an imagej plugin to measure the surface tension from an image of a pendent drop, J. Open Res. Softw. 4(1) (2016) e3-e3.

# Appendix

**Derivation of the drainage hydrodynamic force:**

For small Reynolds number and small separation distances $d < \sqrt{2Rd}$, the fluid flow between the sphere and the surface can be analyzed using lubrication theory. The key assumptions are that the fluid flow is slow (low Reynolds number), and the flow is confined in the thin gap between the sphere and the surface.

The governing equations are:

the velocity gradient in the $z$-direction:

$$\frac{\partial v_z}{\partial z} = -\frac{1}{r}\frac{\partial}{\partial r}(rv_r) \qquad (A1)$$

the pressure gradient in the radial direction:

$$\frac{\partial p}{\partial r} = \eta \frac{\partial^2 v_r}{\partial z^2} \qquad (A2)$$

the pressure gradient is constant in the vertical direction:

$$\frac{\partial p}{\partial z} = 0 \qquad (A3)$$

here $\eta$ is the dynamic viscosity of the fluid, $v_r$ and $v_z$ represent the radial and vertical components of the velocity, respectively.

The thickness of the confined liquid between the sphere and the flat surface is approximated as:

$$h(r) = d + \frac{r^2}{2R} \qquad (A4)$$

where $d$ is the distance between the spherical particle and the flat surface, and $R$ is the radius of curvature of the sphere.

For hydrophilic surfaces, the non-slip boundary conditions are assumed, which are:

on the flat surface at $z = 0$:

$$v_r(z=0) = 0, v_z(z=0) = 0 \qquad (A5)$$

on the surface of the sphere at $z = h$:

$$v_r(z=h) = 0, v_z(z=h) = -V \qquad (A6)$$

These boundary conditions reflect the no-slip behavior of the fluid at both surfaces: the fluid velocity at the solid surfaces matches the velocity of the solid boundaries.

By solving the set of **Eqs. (A1), (A2)** and **(A3)** and using the boundary conditions **Eqs. (A5)** and **(A6)** we obtain the lubrication equation:

$$V = \frac{1}{12\eta r}\frac{\partial}{\partial r}\left(rh^3 \frac{\partial p}{\partial r}\right) \qquad (A7)$$

By integration, we get:

$$\frac{\partial p}{\partial r} = \frac{6\eta r}{h^3} V \qquad (A8)$$

**Equation. (A8)** gives the expression of the pressure gradient as a function of the velocity of the sphere and the thickness profile of the liquid layer. To calculate the hydrodynamic force $F_h$ between the sphere and the surface, we use the relation:

$$F_h = 2\pi \int_0^\infty rp(r)\, dr = -\pi \int_0^\infty r^2 \frac{\partial p}{\partial r}\, dr \qquad (A9)$$

By substituting **Eq.(A8)** in **Eq.(A9)** we arrive at the following equation for the hydrodynamic force:

$$F_h = -\frac{6\pi \eta R^2}{d} V$$